\documentclass[12pt]{article}
\usepackage{graphicx}
\usepackage{caption}
\usepackage{subcaption}
\usepackage{epstopdf}
\usepackage{epsfig}
\usepackage{amsmath}
\usepackage{amssymb}
\usepackage{multirow}
\usepackage{rotating}
\usepackage{array}
\usepackage{amsfonts}
\usepackage{afterpage}
\def\lsim{\mathrel{\raise.3ex\hbox{$<$\kern-.75em\lower 1ex\hbox{$\sim$}}}}
\def\gsim{\mathrel{\raise.3ex\hbox{$>$\kern-.75em\lower 1ex\hbox{$\sim$}}}}

\def\be{\begin{equation}}
\def\ee{\end{equation}}
\def\bea{\begin{eqnarray*}}
\def\eea{\end{eqnarray*}}

\begin{document}
\title{Generalized perturbations in neutrino mixing}
\author{Jiajun Liao$^{1,3}$, D. Marfatia$^{1,3}$, and K. Whisnant$^{2,3}$\\
\\
\small\it $^1$Department of Physics and Astronomy, University of Hawaii at Manoa, Honolulu, Hawaii 96822, USA\\
\small\it $^2$Department of Physics and Astronomy, Iowa State University, Ames, Iowa 50011, USA\\
\small\it $^3$Kavli Institute for Theoretical Physics, University of California, Santa Barbara, California 93106, USA}
\date{}
\maketitle

\begin{abstract}
We derive expressions for the
neutrino mixing parameters that result from complex perturbations on
(1) the Majorana neutrino mass matrix (in the basis of charged lepton
mass eigenstates) and on (2) the charged lepton mass matrix, for
arbitrary initial (unperturbed) mixing matrices. In the first case, we
find that the phases of the elements of the perturbation matrix, and
the initial values of the Dirac and Majorana phases, strongly impact
the leading-order corrections to the neutrino mixing parameters and phases. For
experimentally compatible scenarios wherein the initial neutrino mass
matrix has $\mu-\tau$ symmetry, we find that the Dirac phase can take
any value under small perturbations. Similarly, in the second case,
perturbations to the charged lepton mass matrix can generate large
corrections to the mixing angles and phases of the Pontecorvo-Maki-Nakagawa-Sakata (PMNS) matrix.  As an
illustration of our generalized procedure, we apply it to a situation
in which nonstandard scalar and nonstandard vector interactions
simultaneously affect neutrino oscillations.
\end{abstract}

\newpage

\section{Introduction}

After decades of neutrino oscillation experiments, the
mixing pattern in the lepton sector has been well
established~\cite{Agashe:2014kda}. There are one
small and two large mixing angles, and
two mass-squared differences that differ by a factor of 30 in the neutrino sector. Numerous
neutrino mixing scenarios have been proposed in the literature to
explain such a nontrivial mixing pattern; for a recent review, see
Ref.~\cite{King:2013eh}. The most attractive scenarios are those with
mixing patterns motivated by simple symmetries, such as tri-bimaximal
mixing~\cite{TBM}, bimaximal mixing~\cite{BM}, and golden
ratio mixing~\cite{GR}. All three mixing scenarios have
\(\theta_{13}=0\), \(\theta_{23}=45^\circ\), and are a subset of the
more general \(\mu-\tau\) symmetry~\cite{Lam:2001fb,Liao:2012xm}. However, recent measurements of
$\theta_{13}$ by short-baseline reactor experiments Daya
Bay~\cite{An:2012eh}, RENO~\cite{Ahn:2012nd}, Double Chooz~\cite{Abe:2013sxa},
and long-baseline accelerator experiments T2K~\cite{T2K},
MINOS~\cite{MINOS} strongly disfavor \(\theta_{13}=0\). Therefore,
models with simple symmetries need additional features to explain the
observed neutrino mixing pattern.

A modified approach to explain the data is to treat the simple mixing scenarios as the
underlying model and add perturbations to accommodate the
discrepancy between theoretical predictions and experimental
data. In Ref.~\cite{Liao:2012xm}, we took $\mu-\tau$ symmetry
 (in the charged lepton basis), as the underlying model,
and added real perturbations to the Majorana neutrino mass matrices to
explain the data. We found that small perturbations can cause large
corrections to $\theta_{12}$, and the experimental data can be
explained by most $\mu-\tau$ symmetric mixing scenarios with
perturbations of similar magnitude.

After the discovery that the mixing angle $\theta_{13}$ is relatively large,
many scenarios without \(\mu-\tau\) symmetry have been proposed~\cite{Toorop:2011jn}. 
Motivated by this, and by the fact that real perturbations
have no effect on the Dirac and Majorana phases, in this
paper we consider arbitrary initial mixing for the underlying model
and generalize the perturbation results to the complex space. Under the assumption that the
charged lepton mass matrix is diagonal, we
derive analytic formulas for the leading-order (LO) corrections to the
three mixing angles and the Dirac and Majorana phases. We find that the
phases of the elements of the perturbation matrix, and the initial
values of the Dirac and Majorana phases, strongly impact the LO
corrections to the neutrino mixing parameters. We also perform a
numerical study of complex perturbations on initial neutrino mass
matrices with $\mu-\tau$ symmetry. We find that the Dirac phase can
take any value under small perturbations for experimentally compatible
scenarios.


Since the mixings in both the charged lepton sector and neutrino sector
contribute to the observed Pontecorvo-Maki-Nakagawa-Sakata (PMNS) matrix, we explore the case in which the
charged lepton mass matrix is not diagonal, and consider small complex
perturbations to the charged lepton mass matrix as well. Since the
initial mixing matrix in the charged lepton sector is unconstrained,
small perturbations in the charged lepton sector could have large
effects on the 1-2 mixing in the charged lepton sector, which lead to large 
changes in the mixing angles and phases in the PMNS matrix.

In addition, as an application of our generalized perturbation
procedure, we study neutrino oscillations with matter effects from
both nonstandard scalar and nonstandard vector interactions. 
Nonstandard scalar interactions add small
perturbations to the neutrino mass matrix, which yield corrections to
the vacuum mixing angles and mass-squared differences. By using our
formalism, we demonstrate how expressions for neutrino oscillation
probabilities that simultaneously depend on nonstandard scalar and nonstandard vector interactions, can
be obtained.

This paper is organized as follows. In Section~2, we work in the diagonal charged lepton basis
and derive analytic formulas for the mixing parameters that result from real and complex perturbations.
 In Section~3, we perform a
numerical analysis of complex perturbations to neutrino mass
matrices with $\mu-\tau$ symmetry. In Section~4, we calculate 
corrections to the three mixing angles and phases in the PMNS matrix from
perturbations in the charged lepton sector. In Section~5, we apply our
perturbation results to study neutrino oscillations with both nonstandard
scalar and nonstandard vector interactions. We summarize our results
in Section~6.

\section{Perturbations on the neutrino mass matrix}
Our goal in this section is to obtain the LO corrections to all
the physical parameters under small perturbations on the
initial Majorana neutrino mass matrix assuming the charged lepton mass matrix to be diagonal. The final (resultant) mass
matrix can be written as the sum of an initial matrix $M_0$ and a
perturbation matrix $E$, i.e.,
\begin{align}
M=M_0+E=U_0^*\overline{M_0}U_0^\dagger+\begin{pmatrix}
   \epsilon_{11} & \epsilon_{12} & \epsilon_{13} \\
   \epsilon_{12} & \epsilon_{22} & \epsilon_{23} \\
   \epsilon_{13} & \epsilon_{23} & \epsilon_{33}
   \end{pmatrix}\,,
\label{eq:perturbation}
\end{align}
where $\overline{M_0}=\text{diag}(m_1^0,m_2^0,m_3^0)$, and $U_0$ is
the initial mixing matrix.

The final mass matrix can also be written as
\begin{equation}
M=U^*\overline{M}U^\dagger,
\end{equation}
where $U$ and $\overline{M}$ have the same form as $U_0$ and
$\overline{M_0}$. From neutrino oscillation
experiments we know that $m_1$ and $m_2$ are nearly degenerate, so
here we assume $ |\epsilon_{ij}|,
|\delta m_{21}^0| \ll|\delta m_{31}^0|$.

\subsection{Real case}
For simplicity, we consider the real case first. The mixing matrix
$U_0$ can be written as
\begin{align}
U_0=R_{23}^0R_{13}^0R_{12}^0\,,
\end{align}
where $R_{ij}^0$ is the rotation matrix in the $i-j$ plane with a
rotation angle $\theta_{ij}^0$. Then Eq.~(\ref{eq:perturbation}) can
be rewritten as
\begin{align}
M &= m_1^0\text{I} + R_{23}^0R_{13}^0R_{12}^0\begin{pmatrix}
   0 & 0 & 0 \\
   0 & \delta m_{21}^0 & 0 \\
   0 & 0 & \delta m_{31}^0
   \end{pmatrix}(R_{12}^0)^T(R_{13}^0)^T(R_{23}^0)^T+E \nonumber\\
  &= m_1^0\text{I} + R_{23}^0R_{13}^0\left[R_{12}^0\begin{pmatrix}
   0 & 0 & 0 \\
   0 & \delta m_{21}^0 & 0 \\
   0 & 0 & \delta m_{31}^0
   \end{pmatrix}(R_{12}^0)^T+E'\right](R_{13}^0)^T(R_{23}^0)^T\,,
\label{eq:squarebracket}
\end{align}
where $E'=(R_{13}^0)^T(R_{23}^0)^T E R_{23}^0R_{13}^0$, $\delta
m_{ji}^0=m_j^0-m_i^0$, and $I$ is the $3\times 3$ identity matrix. We
employ the following notation:
\begin{align}
\epsilon_1&=\epsilon_{11}\,,\quad
\epsilon_2=\epsilon_{12}c_{23}^0-\epsilon_{13}s_{23}^0\,,\quad
\epsilon_3=\epsilon_{12}s_{23}^0+\epsilon_{13}c_{23}^0\,,\nonumber \\
\epsilon_4&=\epsilon_{22}(c_{23}^0)^2+\epsilon_{33}(s_{23}^0)^2-\epsilon_{23}s_{2\times 23}^0\,, \nonumber \\
\epsilon_5&=\epsilon_{23}c_{2\times 23}^0+\frac{1}{2}(\epsilon_{22}-\epsilon_{33})s_{2\times 23}^0\,,\nonumber \\
\epsilon_6&=\epsilon_{22}(s_{23}^0)^2+\epsilon_{33}(c_{23}^0)^2+\epsilon_{23}s_{2\times 23}^0\,,
\end{align}
where $c_{ij}^0$, $s_{ij}^0$, $c_{2\times ij}^0$ and $s_{2\times ij}^0$ denote $\cos\theta_{ij}^0$, $\sin\theta_{ij}^0$, $\cos(2\theta_{ij}^0)$ and $\sin(2\theta_{ij}^0)$, respectively. Then $\epsilon'_{ij}\equiv(E')_{ij}$ can be written explicitly as
\begin{align}
\epsilon'_{11}&=\epsilon_1(c_{13}^0)^2+\epsilon_6(s_{13}^0)^2-\epsilon_3s_{2\times 13}^0\,,\quad
\epsilon'_{12}=\epsilon_2c_{13}^0-\epsilon_5s_{13}^0\,,\nonumber\\
\epsilon'_{13}&=\epsilon_3c_{2\times 13}^0+\frac{1}{2}(\epsilon_1-\epsilon_6)s_{2\times 13}^0\,,\quad
\epsilon'_{22}=\epsilon_4\,,\nonumber\\
\epsilon'_{23}&=\epsilon_2s_{13}^0+\epsilon_5c_{13}^0\,,\quad
\epsilon'_{33}=\epsilon_1(s_{13}^0)^2+\epsilon_6(c_{13}^0)^2+\epsilon_3s_{2\times 13}^0\,.
\end{align}
In order to obtain the final mixing matrix that diagonalizes $M$, we
use a procedure that is similar to that in
Ref.~\cite{King:2002nf}. We first put zeros in the 2-3 and 1-3 entries
of the matrix in the square bracket of Eq.~(\ref{eq:squarebracket}) by
using rotations $R_{23}(\delta'_{23})$ and $R_{13}(\delta'_{13})$,
respectively. To LO in $\mathcal{O}(|\epsilon_{ij}|/|\delta
m_{31}^0|)$, we have
\begin{align}
\delta'_{23}&\approx\frac{\epsilon'_{23}}{\delta m_{31}^0}\,,\quad
\delta'_{13}\approx\frac{\epsilon'_{13}}{\delta m_{31}^0}\,,
\end{align} 
and the LO correction to $m_3^0$ is
\begin{equation}
\delta m_3=\epsilon'_{33}\,.
\end{equation}
Note that since $ |\epsilon_{ij}|\ \ll|\delta m_{31}^0|$,
after the two rotations in the 2-3 and 1-3 planes, the matrix in the
square bracket of Eq.~(\ref{eq:squarebracket}) becomes block diagonal
and the 1-2 submatrix remains unchanged to leading order. Hence we can
rewrite Eq.~(\ref{eq:squarebracket}) as
\begin{align}
M = m_1^0\text{I} + V\begin{pmatrix}
M' & 0 \\
0& \delta m_{31}^0
\end{pmatrix}V^T+O(|\epsilon_{ij}|^2/\delta m_{31}^0)\,,
\end{align}
where $V=R_{23}^0R_{13}^0R_{23}(\delta'_{23})R_{13}(\delta'_{13})R_{12}^0$, and 
\begin{align}
M'=\begin{pmatrix}
   \epsilon'_{11}(c_{12}^0)^2+\epsilon'_{22}(s_{12}^0)^2-\epsilon'_{12}s_{2\times 12}^0 & \epsilon'_{12}c_{2\times 12}^0+\frac{1}{2}(\epsilon'_{11}-\epsilon'_{22})s_{2\times 12}^0  \\
   \epsilon'_{12}c_{2\times 12}^0+\frac{1}{2}(\epsilon'_{11}-\epsilon'_{22})s_{2\times 12}^0 & \epsilon'_{11}(s_{12}^0)^2+\epsilon'_{22}(c_{12}^0)^2+\epsilon'_{12}s_{2\times 12}^0+\delta m_{21}^0 
   \end{pmatrix}\,,
\end{align}
which can be diagonalized by the rotation $R_{12}(\xi')$ with
\begin{equation}
\xi'=\frac{1}{2}\arctan\frac{2\epsilon'_{12}c_{2\times 12}^0-(\epsilon'_{22}-\epsilon'_{11})s_{2\times 12}^0}{(\epsilon'_{22}-\epsilon'_{11})c_{2\times 12}^0+2\epsilon'_{12}s_{2\times 12}^0+\delta m_{21}^0}\,.
\end{equation}
The corrections to $m_1$ and $m_2$ can be written as
\begin{align}
\delta m_{i}=\frac{\epsilon'_{11}+\epsilon'_{22}}{2}\pm\frac{1}{2}\left[\delta m_{21}^0-\sqrt{\Delta}\right]\,,
\end{align}
where $\Delta=
(\delta m_{21}^0)^2 + 4(\epsilon'_{12})^2 + (\epsilon'_{22}-\epsilon'_{11})^2
+ 2\delta m_{21}^0\left[2\epsilon'_{12}s_{2\times 12}^0
+ (\epsilon'_{22} - \epsilon'_{11})c_{2\times 12}^0\right]$, and the
plus (minus) sign is for $i=1$ ($2$). The final mass matrix is
diagonalized by the following mixing matrix
\begin{equation}
U=R_{23}^0R_{13}^0R_{23}(\delta'_{23})R_{13}(\delta'_{13})R_{12}^0R_{12}(\xi')\,.
\end{equation}
By comparing it to the standard parametrization, we find
the LO corrections to the three mixing angles to be
\begin{align}
\delta\theta_{13}&=\delta'_{13}=\frac{\epsilon'_{13}}{\delta m_{31}^0}\,,\nonumber\\
\delta\theta_{23}&=\frac{\delta'_{23}}{c_{13}^0}=\frac{\epsilon'_{23}}{c_{13}^0\delta m_{31}^0}\,,\nonumber\\
\delta\theta_{12}&=\xi'=\frac{1}{2}\arctan\frac{2\epsilon'_{12}c_{2\times 12}^0-(\epsilon'_{22}-\epsilon'_{11})s_{2\times 12}^0}{(\epsilon'_{22}-\epsilon'_{11})c_{2\times 12}^0+2\epsilon'_{12}s_{2\times 12}^0+\delta m_{21}^0}\,,
\label{eq:realdth}
\end{align} 
where we have ignored the next-to-leading-order correction to
$\theta_{12}$, which is $\mathcal{O}(|\epsilon_{ij}|/|\delta
m_{31}^0|)$. For $\theta_{13}^0=0$ and $\theta_{23}^0=\pi/4$, it is
easy to verify that the corrections in Eq.~(\ref{eq:realdth}) yield
the results of Ref.~\cite{Liao:2012xm} for the LO
corrections,\footnote{Also, for the next-to-leading-order correction to
$\theta_{12}$, we obtain Eq.~(14) of
Ref.~\cite{Liao:2012xm}, except that $\delta m_{21}^0$ in the denominator
should be replaced by $\delta m_{21}^{(1)}\equiv(m_2^0+\delta
m_2)-(m_1^0+\delta m_1)$.} which were obtained using degenerate
perturbation theory. As noted in Ref.~\cite{Liao:2012xm}, the near degeneracy
of $m_1$ and $m_2$ ($|\delta m_{21}^0| \ll |\delta m_{31}^0|$) implies
that $\delta \theta_{12}$ can be large for small perturbations
($|\epsilon_{ij}| \ll |\delta m_{31}^0|$). 

\subsection{Complex case}
\label{set2.2}
For the complex case, the most general form for $U_0$ is
\begin{equation}
U_0=R_{23}(\theta_{23}^0)U_{13}(\theta_{13}^0,\delta^0)R_{12}(\theta_{12}^0)P(\phi_2^0,\phi_3^0)\,,
\label{eq:U0}
\end{equation}
where
\begin{align}
R_{23}(\theta_{23}^0)&=\begin{pmatrix}
1 & 0 & 0\\
0 & c_{23}^0 & s_{23}^0 \\
0 & -s_{23}^0 & c_{23}^0
\end{pmatrix}, \quad U_{13}(\theta_{13}^0,\delta^0)=\begin{pmatrix}
c_{13}^0 & 0 & e^{-i\delta^0}s_{13}^0 \\
0 & 1 & 0 \\
-e^{i\delta^0}s_{13}^0 & 0 & c_{13}^0
\end{pmatrix}\,, \nonumber \\
R_{12}(\theta_{12}^0)&=\begin{pmatrix}
c_{12}^0 & s_{12}^0 & 0 \\
-s_{12}^0 & c_{12}^0 & 0 \\
0 & 0 & 1
\end{pmatrix}, \quad P(\phi_2^0,\phi_3^0)=\begin{pmatrix}
   1 & 0 & 0 \\[0.3em]
   0 & e^{i\phi_2^0/2} & 0 \\[0.3em]
   0 & 0 & e^{i\phi_3^0/2}
   \end{pmatrix}\,.
\end{align}
Because of the nonzero Majorana phases, in general, the mixing
matrix would not remain unchanged if we subtract the identity matrix
multiplied by a constant from the mass matrix. Hence we use a
slightly different procedure to obtain the LO corrections for the
complex case. We rewrite the final mass matrix as
%
\begin{align}
M=U_0^*\overline{M_0}U_0^\dagger+E =U_0^*\left[\overline{M_0}+\tilde{E}\right]U_0^\dagger\,,
\end{align}
where $\tilde{E}=U_0^T E U_0$  can be explicitly written as
\begin{align}
\tilde{E}=\begin{pmatrix}
a & be^{i\phi_2^0/2} & de^{i\phi_3^0/2} \\
be^{i\phi_2^0/2} & ce^{i\phi_2^0} & f e^{i(\phi_2^0+\phi_3^0)/2} \\
de^{i\phi_3^0/2} & f e^{i(\phi_2^0+\phi_3^0)/2} & g e^{i\phi_3^0}\\
\end{pmatrix}\,,
\end{align}
with
\begin{align}
a&=\epsilon_4(s_{12}^0)^2+[\epsilon_1(c_{13}^0)^2-\epsilon_3s_{2\times 13}^0 e^{i\delta^0}+ \epsilon_6(s_{13}^0)^2e^{2i\delta^0}](c_{12}^0)^2+(\epsilon_5s_{13}^0e^{i\delta^0}-\epsilon_2c_{13}^0)s_{2\times 12}^0\,,\nonumber \\
b&=\epsilon_2c_{13}^0c_{2\times 12}^0+[\epsilon_1(c_{13}^0)^2-\epsilon_4+\epsilon_6(s_{13}^0)^2e^{2i\delta^0}]c_{12}^0s_{12}^0 -[\epsilon_3s_{2\times 12}^0s_{13}^0c_{13}^0+\epsilon_5c_{2\times 12}^0s_{13}^0]e^{i\delta^0}\,,\nonumber \\
c&=\epsilon_4(c_{12}^0)^2+[\epsilon_1(c_{13}^0)^2-\epsilon_3s_{2\times 13}^0e^{i\delta^0}+ \epsilon_6(s_{13}^0)^2e^{2i\delta^0}](s_{12}^0)^2-(\epsilon_5s_{13}^0e^{i\delta^0}-\epsilon_2c_{13}^0)s_{2\times 12}^0\,,\nonumber\\
d&=(\epsilon_1c_{12}^0c_{13}^0-\epsilon_2s_{12}^0)s_{13}^0e^{-i\delta^0}-\epsilon_6c_{12}^0c_{13}^0s_{13}^0e^{i\delta^0}+\epsilon_3c_{12}^0c_{2\times 13}^0-\epsilon_5c_{13}^0s_{12}^0\,,\nonumber\\
f&=(\epsilon_1s_{12}^0c_{13}^0+\epsilon_2c_{12}^0)s_{13}^0e^{-i\delta^0}-\epsilon_6s_{12}^0c_{13}^0s_{13}^0e^{i\delta^0}+\epsilon_3s_{12}^0c_{2\times 13}^0+\epsilon_5c_{13}^0c_{12}^0\,,\nonumber \\
g&=\epsilon_6(c_{13}^0)^2+\epsilon_3s_{2\times 13}^0e^{-i\delta^0}+\epsilon_1(s_{13}^0)^2e^{-2i\delta^0}\,,
\label{eq:abc}
\end{align}
%

Similar to the real case, we apply a unitary matrix $U_\delta$ to
$N\equiv\overline{M_0}+\tilde{E}$ such that there are zeros in the 2-3
and 1-3 entries of the matrix $U_\delta^T N U_\delta$. Since
$ |\epsilon_{ij}| \ll|\delta m_{31}^0|$, to LO in
$\mathcal{O}(|\epsilon_{ij}|/|\delta m_{31}^0|)$, $U_\delta$ can be
written as
\begin{equation}
U_\delta=\begin{pmatrix}
1 & 0 & \delta_{13} \\
0 & 1 & \delta_{23} \\
-\delta_{13}^{*} & -\delta_{23}^{*} & 1\\
\end{pmatrix}\,,
\end{equation}
where
\begin{align}
\delta_{13}\approx\frac{|d|e^{-i\phi_{13}}}{|m_3^0-m_1^0e^{-2i\phi_{13}}|}\,,\quad \delta_{23}\approx\frac{|f| e^{-i\phi_{23}}}{|m_3^0-m_1^0e^{-2i\phi_{23}}|}\,,
\label{eq:delta1323}
\end{align}
with $\tan\phi_{13}=\frac{m_3^0+m_1^0}{m_3^0-m_1^0}\tan
\left[\arg(d)+\phi_3^0/2\right]$ and $\tan\phi_{23}=
\frac{m_3^0+m_1^0}{m_3^0-m_1^0}\tan
\left[\arg(f)+\frac{\phi_2^0+\phi_3^0}{2}\right]$. After
block-diagonalization, the LO correction to $m_3$ is
\begin{align}
\delta m_3=\left|m_3^0+g e^{i\phi_3^0}\right|-m_3^0\,.
\end{align}

Note that the 1-2 submatrix of $N$ remains unchanged to leading order
after the block-diagonalization. Using the procedure described in 
Appendix~\ref{ap:diagonalization}, we diagonalize this submatrix using the unitary matrix
\begin{align}
U_{12}(\xi,\phi) =
\begin{pmatrix}
c_\xi & s_\xi e^{-i\phi} & 0 \\
-s_\xi e^{i\phi} & c_\xi & 0\\
0 & 0 & 1 \\
\end{pmatrix}\,,
\end{align}
where
\begin{align}
\phi=\arctan \frac{|a+m_1^0| \sin(\phi_a-\phi_b)-|ce^{i\phi_2^0}+m_2^0|\sin(\phi_c-\phi_b)}{|a+m_1^0|\cos(\phi_a-\phi_b)+|ce^{i\phi_2^0}+m_2^0|\cos(\phi_c-\phi_b)}\,,
\label{eq:phi}
\end{align}
\begin{align}
\xi=\frac{1}{2}\arctan\frac{2|b|}{|ce^{i\phi_2^0}+m_2^0|\cos(\phi_c+\phi-\phi_b)-|a+m_1^0|\cos(\phi_a-\phi-\phi_b)}\,,
\label{eq:xi}
\end{align}
with $\phi_a=\arg(a+m_1^0)$, $\phi_b=\arg(b)+\phi_2^0/2$ and
$\phi_c=\arg(ce^{i\phi_2^0}+m_2^0)$. In addition, we obtain the LO
corrections to $m_1$ and $m_2$ as
\begin{align}
\delta m_1&=\left|(a+m_1^0) c_\xi^2+(c e^{i\phi_2^0}+m_2^0) s_\xi^2e^{2i\phi}-2bs_\xi c_\xi e^{i\phi}\right|-m_1^0 \,,\nonumber \\
\delta m_2&=\left|(a+m_1^0) s_\xi^2e^{-2i\phi}+(c e^{i\phi_2^0}+m_2^0) c_\xi^2+2b s_\xi c_\xi e^{-i\phi}\right|-m_2^0\,.
\end{align}

The final mixing matrix that diagonalizes $M$ and makes the diagonal
elements real and non-negative can be written as
\begin{align}
U=U_0U_\delta U_{12}(\xi,\phi)P\,,
\end{align}
where $P=\text{diag}(e^{i\omega_1/2},e^{i\omega_2/2},e^{i\omega_3/2})$, and
\begin{align}
\omega_1&=-\arg \left[(a+m_1^0) c_\xi^2+(c e^{i\phi_2^0}+m_2^0) s_\xi^2e^{2i\phi}-2bs_\xi c_\xi e^{i\phi}\right] \,,\nonumber \\
\omega_2&=-\arg \left[(a+m_1^0) s_\xi^2e^{-2i\phi}+(c e^{i\phi_2^0}+m_2^0) c_\xi^2+2b s_\xi c_\xi e^{-i\phi}\right]\,, \nonumber \\
\omega_3&=-\arg\left(m_3^0+g e^{i\phi_3^0}\right)\,.
\end{align}
%

As shown in Appendix~\ref{ap:rtmulti}, the
right-multiplication of $U_{12}(\xi,\phi)$ does not change
$\theta_{13}$ and $\theta_{23}$. Hence, the LO corrections to
$\theta_{13}$ and $\theta_{23}$ come from the right-multiplication of
$U_\delta$. Since $\delta_{13}$ and $\delta_{23}$ are suppressed by a
factor of $|\epsilon_{ij}|/|\delta m_{31}^0|$, while $\xi$ and $\phi$
are not, the LO corrections to $\theta_{12}$ and the Dirac phases come
from the right-multiplication of $U_{12}(\xi,\phi)$, and the LO
corrections to the Majorana phases come from both $U_{12}(\xi,\phi)$
and $P$.

By comparing $U$ to the standard parametrization, we obtain the LO
corrections to the three mixing angles:
\begin{align}
\delta\theta_{13} &=
\frac{|d|c_{12}^0\cos(\delta^0 - \frac{\phi_3^0}{2} - \phi_{13})}
{|m_3^0-m_1^0e^{-2i\phi_{13}}|} +\frac
{|f|s_{12}^0\cos(\delta^0+\frac{\phi_2^0-\phi_3^0}{2}-\phi_{23})}
{|m_3^0-m_1^0e^{-2i\phi_{23}}|}
\label{eq:dth13}
\end{align}
\begin{align}
\delta\theta_{23} &=
-\frac{|d|s_{12}^0\cos(\frac{\phi_3^0}{2}+\phi_{13})}
{|m_3^0-m_1^0e^{-2i\phi_{13}}|} +\frac
{|f|c_{12}^0\cos(\frac{\phi_2^0-\phi_3^0}{2}-\phi_{23})}
{|m_3^0-m_1^0e^{-2i\phi_{23}}|}
\label{eq:dth23}
\end{align}
\begin{align}
\delta\theta_{12} &= \arcsin \sqrt{\sin^2(\theta_{12}^0+\xi)
-\sin(2\theta_{12}^0)\sin(2\xi)\sin^2\frac{\phi_2^0+2\phi}{4}}
-\theta_{12}^0\,,
\label{eq:dth12}
\end{align}
where $t_{ij}^0$ denotes $\tan\theta_{ij}^0$. The LO corrections to the three phases can be written as 
\begin{equation}
\Delta\delta = \alpha - \beta\,,
\end{equation}
\begin{align}
\Delta\phi_2 &= -2(\alpha + \beta)+\omega_2-\omega_1\,,
\end{align}
\begin{align}
\Delta\phi_3 &= -2\beta+\omega_3-\omega_1\,.
\end{align}
where 
\begin{align}
\alpha &= -\arctan\frac{\tan\theta_{12}^0\tan\xi\sin(\phi_2^0/2+\phi)}
               {1-\tan\theta_{12}^0\tan\xi\cos(\phi_2^0/2+\phi)}\,,
\end{align}
and 
\begin{align}
\beta &= \arctan\frac{\tan\xi\sin(\phi_2^0/2+\phi)}
              {\tan\theta_{12}^0+\tan\xi\cos(\phi_2^0/2+\phi)}\,.
\end{align}


From Eq.~(\ref{eq:dth12}), we see that $\delta\theta_{12}$ varies
from $-\xi$ to $+\xi$ depending on the initial Majorana phase
$\phi_2^0$ and the perturbation phase $\phi$. Since $\xi$ and $\phi$
depend only on the ratios of linear combinations of $\epsilon_{ij}$'s
and $\delta m_{21}^0$, large corrections to $\theta_{12}$ and the Dirac and Majorana
phases are possible even for small perturbations. However, corrections can be small in special cases, e.g., if $\phi_2^0$ is close to $180^\circ$ for the inverted hierarchy, $\phi$ approaches $90^\circ$ and $\xi$ is suppressed by a factor of $|\epsilon_{ij}|/(m_2^0+m_1^0)$, so that the corrections to $\theta_{12}$ and the Dirac and Majorana phases are also small.

Note that the corrections in the
complex case are strongly dependent on the phases of $\epsilon_{ij}$,
and the initial values of the Dirac and Majorana phases. If we take
$\epsilon_{ij}$'s to be real, and set $\delta^0=\phi_2^0=\phi_3^0=0$
in Eqs.~(\ref{eq:dth13}), (\ref{eq:dth23}) and (\ref{eq:dth12}), we recover
Eq.~(\ref{eq:realdth}).

\section{Perturbations to $\mu-\tau$ symmetry}

As an illustration of our analytic results, we study perturbations on initial neutrino mass matrices with $\mu-\tau$ symmetry. There are four classes of mixing with $\mu-\tau$ symmetry~\cite{Liao:2012xm}:
(a) $\theta_{23}^0=45^\circ, \theta_{13}^0=0$; (b)
$\theta_{23}^0=45^\circ, \theta_{12}^0=0$; (c) $\theta_{23}^0=45^\circ,
\theta_{12}^0=90^\circ$; (d) $\theta_{23}^0=45^\circ, \delta^0=\pm 90^\circ$. In Ref.~\cite{Babu:2002dz}, it was shown that the initial class~(a) can be perturbed to 
class~(d) for a specific model. Here we reproduce the results of Ref.~\cite{Babu:2002dz} by applying our general perturbation formulas. 
The complex neutrino mass matrix of Ref.~\cite{Babu:2002dz} can be written (in our phase convention) as
\begin{align}
M=m_0 \begin{pmatrix}
   1+2\delta''' & 0 & 0 \\
   0 & \delta''' & -(1+\delta''') \\
   0 & -(1+\delta''') & \delta'''
   \end{pmatrix}+m_0 \begin{pmatrix}
      2\delta' & \delta'' & -\delta''^* \\
      \delta'' & 0 & 0 \\
      -\delta''^* & 0 & 0
      \end{pmatrix}\,,
\label{eq:Mex}
\end{align}
where $m_0$ is a common mass parameter, $\delta'''$, $\delta'$ are real and $|\delta'|$, $|\delta''|\ll|\delta'''|$. We treat the first term on the right-hand side of Eq.~(\ref{eq:Mex}) as the initial mass matrix and the second term as the perturbation. 
 The initial mass matrix has class (a) $\mu-\tau$ symmetry. In the standard parametrization, we have $\theta_{23}^0=\frac{\pi}{4}$, $\theta_{12}^0=\theta_{13}^0=\phi_2^0=0$ and $\phi_3^0=\pi$. The three initial masses are $m_1^0=m_2^0=m_0(1+2\delta''')$, and $m_3^0=m_0$. In this case, Eq.~(\ref{eq:abc}) is greatly simplified:
\begin{align}
a=2m_0\delta'\,,\quad \quad b=\sqrt{2}m_0\text{Re}\left(\delta''\right)\,, \quad\quad
c=f=g=0\,,\quad \quad d=i\sqrt{2}m_0\text{Im}\left(\delta''\right)\,.
\end{align}
From Eqs.~(\ref{eq:delta1323}),~(\ref{eq:phi}) and~(\ref{eq:xi}), we find 
\begin{align}
\delta_{23}=0\,,\quad\quad\delta_{13}\approx\frac{\text{Im}\left(\delta''\right)}{\sqrt{2}\delta'''}\,,\quad\quad
\phi=0\,,\quad\quad\xi=\frac{1}{2}\arctan\frac{\sqrt{2}\text{Re}\left(\delta''\right)}{-\delta'}\,.
\end{align}
Then the final mixing matrix can be written as
\begin{align}
U&=R_{23}(\frac{\pi}{4})P(0,\pi)R_{13}(\delta_{13})R_{12}(\xi)\\\nonumber
&=R_{23}(\frac{\pi}{4})U_{13}(\delta_{13},\frac{\pi}{2})R_{12}(\xi)P(0,\pi)\,.
\end{align}
Hence, the final mixing angles and the Dirac phase are
\begin{align}
\theta_{23}=\frac{\pi}{4}\,,\quad \quad\theta_{12}=\frac{1}{2}\arctan\frac{\sqrt{2}\text{Re}\left(\delta''\right)}{-\delta'}\,,\quad\quad
\theta_{13}=\frac{\text{Im}\left(\delta''\right)}{\sqrt{2}\delta'''}\,,\quad \quad \delta=\frac{\pi}{2}\,,
\end{align}
as in Ref.~\cite{Babu:2002dz}. Note that the initial class (a) is perturbed to class (d), and that the large change of the Dirac phase $\delta$ coincides with the deviation of $\theta_{13}$ from 0. 

The general form of the neutrino mass matrix with class (d) $\mu-\tau$ symmetry and its associated generalized CP symmetry has been recognized in 
Ref.~\cite{Grimus:2003yn}, and deviations from it were discussed in Ref.~\cite{Ma:2015gka}. It has been shown in Ref.~\cite{Grimus:2003yn} that the general forms of the neutrino mass matrices with class (a) and (d) $\mu-\tau$ symmetry are (in our phase convention)
\begin{align}
M_a=\begin{pmatrix}
   x & y & -y \\
   y & z & -w \\
   -y & -w & z
   \end{pmatrix}\,,\quad\text{and}\quad
M_d=\begin{pmatrix}
   u & r & -r^* \\
   r & s & -v \\
   -r^* & -v & s^*
   \end{pmatrix}\,,
\end{align}
respectively. Here $x$, $y$, $z$, $w$, $r$, $s$ are complex and $u$, $v$ are real. Hence, any perturbation matrix of the form
\begin{align}
E=\begin{pmatrix}
   \text{Re}(\epsilon_{11})-i\text{Im}(x) & \epsilon_{12} & -\epsilon_{12}^*+2i\text{Im}(y) \\
   \epsilon_{12} & \epsilon_{22} & \text{Re}(\epsilon_{23})+i\text{Im}(w) \\
   -\epsilon_{12}^*+2i\text{Im}(y) & \text{Re}(\epsilon_{23})+i\text{Im}(w) & \epsilon_{22}^*-2i\text{Im}(z)
   \end{pmatrix}\,,
\end{align}
perturbs the initial mass matrix with class (a) $\mu-\tau$ symmetry to class (d) $\mu-\tau$ symmetry.

We now perform a numerical search to find perturbations that fit the experimental data for
initial neutrino mass matrices with $\mu-\tau$ symmetry. We select class (d) and scan $\theta_{12}^0$ and $\theta_{13}^0$ over the range
$[0, 90^\circ]$. Since the initial mass matrices of classes (a), (b) and~(c) do not depend on $\delta^0$, the perturbation results of class (d) will cover the other classes, e.g., the perturbation results for bimaximal mixing would be the same as that of class (d) with $\theta_{13}^0=0$ and $\theta_{12}^0=45^\circ$.
Since we work in the basis in which the charged lepton mass matrix is
diagonal, the mixing matrix in the neutrino sector is the same
as the observed PMNS matrix. We also choose $m_1=0$ for the normal
hierarchy (or $m_3=0$ for the inverted hierarchy), so the best-fit
values from the global fit in Table~\ref{tab:data} define the
other two final masses and the three final mixing angles.

\begin{table}
\caption{Best-fit values and $2\sigma$ ranges of the oscillation parameters~\cite{Capozzi:2013csa}, with $\delta m^2 \equiv m_2^2-m_1^2$ and $\Delta m^2 \equiv m_3^2-(m_1^2+m_2^2)/2$.}
\begin{center}
\begin{tabular}{|l|*{5}{c|}}\hline
\makebox[4em]{\ \ Parameter}
&\makebox[2em]{$\theta_{12}(^\circ)$}&\makebox[2em]{$\theta_{13}(^\circ)$}&\makebox[2em]{$\theta_{23}(^\circ)$}
&\makebox[6em]{$\delta m^2(10^{-5}\text{eV}^2)$}&\makebox[7em]{$|\Delta m^2|(10^{-3}\text{eV}^2)$}\\\hline
Normal hierarchy&$33.7^{+2.1}_{-2.1}$&$8.80^{+0.73}_{-0.77}$&$41.4^{+6.6}_{-2.6}$&$7.54^{+0.46}_{-0.39}$&$2.43^{+0.12}_{-0.13}$\\\hline
Inverted hierarchy&$33.7^{+2.1}_{-2.1}$&$8.91^{+0.70}_{-0.82}$&$42.4^{+9.5}_{-3.2}$&$7.54^{+0.46}_{-0.39}$&$2.38^{+0.12}_{-0.13}$\\\hline
\end{tabular}
\end{center}
\label{tab:data}
\end{table}

We characterize the size of the perturbation as the root-mean-square (RMS) value of the perturbations,
\begin{equation}
\epsilon_\text{RMS} = \sqrt{\frac{\text{Tr}[E^\dagger E]}{9}}=\sqrt{\frac{\sum_{i,j=1}^3 |\epsilon_{ij}|^2}{9}}\,,
\label{eq:rms}
\end{equation}
where $i$ and $j$ sum over neutrino flavors. $\epsilon_\text{RMS}$ is determined by the three initial masses, two initial Majorana phases, two final Majorana phases and one final Dirac phase. 


\begin{figure}
\centering
\centering
\includegraphics[width=0.7\textwidth]{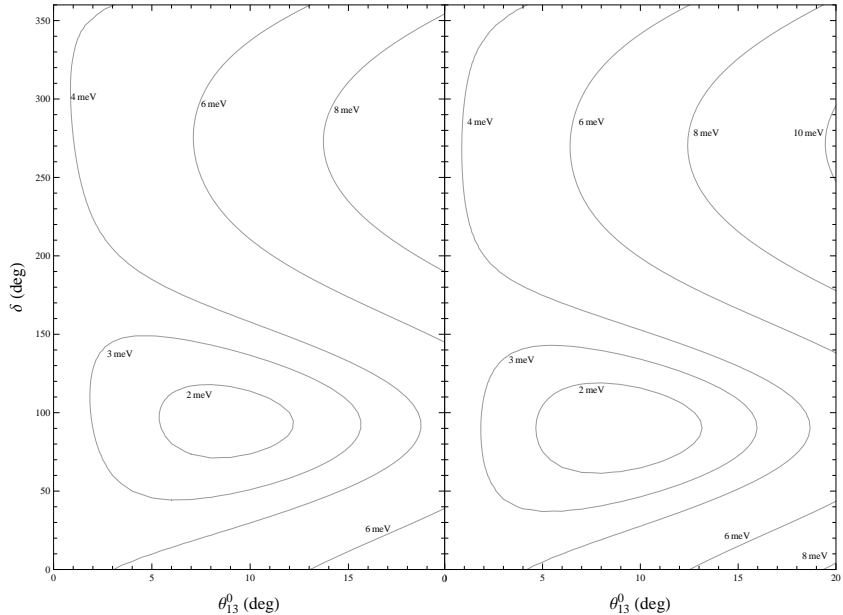}
\caption{Iso-$\epsilon_\text{RMS}^\text{min}$ contours in the
  $(\theta_{13}^0, \delta)$ plane that give the best-fit parameters
  for $\mu-\tau$ symmetry with \(\theta_{23}^0=\theta_{12}^0=45^\circ, \delta^0=
  90^\circ\). The left panel is for the normal hierarchy with $m_1=0$
  and the right panel is for the inverted hierarchy with
  $m_3=0$.}
\label{fg:1}
\end{figure}

The initial Dirac phase in class (d) is fixed to be $\pm 90^\circ$. To
evaluate the change in the Dirac phase due to the perturbations, we
fix $\theta_{12}^0=45^\circ$ and scan over $\delta$ and
$\theta_{13}^0$ to find the minimum RMS value of the perturbation,
$\epsilon_\text{RMS}^\text{min}$, that results in the best-fit
parameters. The results for $\delta^0= 90^\circ$ are shown in
Fig.~\ref{fg:1}. The results for $\delta^0=-90^\circ$ (or
$270^\circ$) are symmetric to those of $\delta^0= 90^\circ$ with
$\delta\rightarrow 360^\circ-\delta$. From Fig.~\ref{fg:1},
we  see that for $\theta_{13}^0 \le 20^\circ$ it is possible for
the final Dirac phase to have any value under small perturbations
($\epsilon_{\text {RMS}}^{\text min} \lsim$~10~meV), i.e., the
correction to the Dirac phase can be large for small perturbations.

\section{Perturbations in the charged lepton sector}
In the basis in which the charged lepton mass matrix is not diagonal, the observed PMNS mixing matrix is
\begin{equation}
U_{PMNS}=U_l^\dagger U_\nu\,,
\end{equation}
where $U_l$ and $U_\nu$ are the mixing matrices in the charged lepton and neutrino sectors, respectively. For an arbitrary charged lepton mass matrix $M_l$, we have 
\begin{align}
(M_l)^\dagger M_l=U_l\overline{M_l}^2(U_l)^\dagger\,,
\end{align}
where $\overline{M_l}=\text{diag}\left(m_e,m_\mu,m_\tau\right)$. 

Suppose the charged lepton mass matrix is also the result of small perturbations to an initial mass matrix, i.e., $M_l=M_l^0+E_l$, where $(E_l)_{ij}\equiv \epsilon_{ij}^l$ and $|\epsilon_{ij}^l|\ll m_\tau$. If the initial mixing matrix in the charged lepton sector is $U_l^0$, i.e.,
\begin{align}
(M_l^0)^\dagger M_l^0=U_l^0(\overline{M_l^0})^2(U_l^0)^\dagger\,,
\end{align}
then to LO, we get
\begin{align}
(M_l)^\dagger M_l&\approx U_l^0(\overline{M_l^0})^2(U_l^0)^\dagger +(M_l^0)^\dagger E_l+E^\dagger M_l^0\nonumber\\
&=U_l^0\left[\overline{M_l^0}^2+ N^l \right](U_l^0)^\dagger\,,
\end{align}
where $N^l=(U_l^0)^\dagger \left[(M_l^0)^\dagger E_l+E^\dagger M_l^0\right]U_l^0$. Note that since $U_l^0$ is unconstrained, the size of each element of the $N^l$ matrix could be of order $m_\tau|\epsilon_{ij}^l|$.

If $(\overline{M_l^0})^2+ N^l$ is diagonalized by a unitary matrix $U_\delta^l$, i.e.,
\begin{align}
\overline{M_l^0}^2+ N^l=\begin{pmatrix}
(m_e^0)^2+N_{11}^l & N_{12}^l & N_{13}^l \\
(N_{11}^l)^* & (m_\mu^0)^2+N_{22}^l & N_{23}^l \\
(N_{13}^l)^* & (N_{23}^l)^* & (m_\tau^0)^2+N_{33}^l
\end{pmatrix}=U_\delta^l\begin{pmatrix}
m_e^2 & 0 & 0 \\
0 & m_\mu^2 & 0 \\
0 & 0 & m_\tau^2
\end{pmatrix}(U_\delta^l)^\dagger\,,
\end{align}
then the PMNS matrix can be written as 
\begin{align}
U_{PMNS}=(U_l^0U_\delta^l)^\dagger U_\nu^0=(U_\delta^l)^\dagger U_0\,,
\label{eq:Upmns}
\end{align}
where $U_0=(U_l^0)^\dagger U_\nu^0$ has the most general form of
Eq.~(\ref{eq:U0}). Since $N_{ij}^l \sim m_\tau|\epsilon_{ij}^l|$, the
2-3 and 1-3 mixing angles in $U_\delta^l$ are suppressed by a factor
of $N_{ij}^l/m_\tau^2\sim |\epsilon_{ij}^l|/m_\tau$. To LO in
$\mathcal{O}(|\epsilon_{ij}^l|/m_\tau)$, $U_\delta^l$ can be parametrized as
\begin{align}
U_\delta^l=\begin{pmatrix}
1 & 0 & 0\\
0 & 1 & \delta_{23}^le^{-i\phi_{23}^l} \\
0 &  -\delta_{23}^le^{i\phi_{23}^l} & 1
\end{pmatrix} \begin{pmatrix}
1 & 0 &  \delta_{13}^le^{-i\phi_{13}^l} \\
0 & 1 & 0 \\
-\delta_{13}^le^{i\phi_{13}^l} & 0 & 1
\end{pmatrix} \begin{pmatrix}
\cos\theta_{12}^l & \sin\theta_{12}^le^{-i\phi_{12}^l} & 0 \\
-\sin\theta_{12}^le^{i\phi_{12}^l} & \cos\theta_{12}^l & 0 \\
0 & 0 & 1
\end{pmatrix}\,,
\end{align}
where 
\begin{align}
\delta_{13}^l &\approx\frac{|N_{13}^l|}{m_\tau^2}\,,\quad \delta_{23}^l \approx \frac{|N_{23}^l|}{m_\tau^2}\,,\nonumber\\
\theta_{12}^l& \approx \frac{1}{2}\arctan\frac{2|N_{12}^l|}{m_\mu^2+N_{22}^l-N_{11}^l} \,,
\label{eq:clcorrection}
\end{align}
and $\phi_{ij}\approx-\arg N_{ij}^l$. 

If $\theta_{12}^l$ is also very small, the LO corrections to the three mixing angles in the PMNS matrix are 
\begin{align}
\delta\theta_{13}&=-\theta_{12}^ls_{23}^0\cos(\delta^0-\phi_{12}^l)-\delta_{13}^lc_{23}^0\cos(\delta^0-\phi_{13}^l)\,,\nonumber \\
\delta\theta_{23}&=-\delta_{23}^l\cos\phi_{23}^l-\delta_{13}^ls_{23}^0t_{13}^0\cos(\delta^0-\phi_{13}^l)+\theta_{12}^lc_{23}^0t_{13}^0\cos(\delta^0-\phi_{12}^l)\,,\nonumber \\
\delta\theta_{12}&=\frac{1}{c_{13}^0}(\delta_{13}^ls_{23}^0\cos\phi_{13}^l-\theta_{12}^lc_{23}^0\cos\phi_{12}^l)\,.
\end{align}

However, in general, since $N_{ij}^l\sim m_\tau|\epsilon_{ij}^l|$, and
if $|\epsilon_{ij}^l|\sim m_\mu^2/m_\tau=6$ MeV, $\theta_{12}^l$ could
be very large, which will give large corrections to the mixing angles
in the PMNS matrix. Thus, the situation in the charged sector is similar to
that in the neutrino sector: the near degeneracy of $m_e$ and $m_\mu$
(on the scale of $m_\tau$) can lead to large corrections in 1-2 space.

For large $\theta_{12}^l$, the analytical
expressions for the corrections to the mixing angles in the PMNS
matrix are cumbersome. Here, as an illustration, we consider the very
simple scenario in which
\begin{align}
U_\delta^l=\begin{pmatrix}
\cos\theta_{12}^l & \sin\theta_{12}^l & 0 \\
-\sin\theta_{12}^l & \cos\theta_{12}^l & 0 \\
0 & 0 & 1
\end{pmatrix}\,.
\label{eq:Udl}
\end{align}
Then from Eq.~(\ref{eq:Upmns}), the final mixing angles in the PMNS matrix are given by
\begin{align}
c_{13}c_{23}=c_{13}^0c_{23}^0\,,
\label{eq:clc23}
\end{align}
\begin{align}
s_{13}^2=(s_{13}^0)^2(c_{12}^l)^2+(s_{23}^0)^2(c_{13}^0)^2(s_{12}^l)^2- 2s_{13}^0c_{13}^0s_{23}^0s_{12}^lc_{12}^l\cos\delta^0\,,
\label{eq:clc13}
\end{align}
\begin{align}
c_{13}^2s_{12}^2&=\left[(c_{12}^lc_{13}^0s_{12}^0-s_{12}^lc_{12}^0c_{23}^0)^2+(s_{12}^l)^2(s_{12}^0)^2(s_{13}^0)^2(s_{23}^0)^2\right.\nonumber\\
&+\left.2s_{12}^ls_{12}^0s_{13}^0s_{23}^0(c_{12}^lc_{13}^0s_{12}^0-s_{12}^lc_{12}^0c_{23}^0)\cos\delta^0\right]\,,
\label{eq:clc12}
\end{align}
where $c_{12}^l$ denotes $\cos\theta_{12}^l$, and $s_{12}^l$ denotes $\sin\theta_{12}^l$. 
As an example, if $\theta_{12}^l$ is the Cabibbo angle and the initial
PMNS matrix has bimaximal symmetry ($\theta_{12}^0 = 45^\circ,
\theta_{13}^0 = 0$), then the resulting $\theta_{12}$ and $\theta_{13}$
are consistent with the observed values to within $2 \sigma$.

\begin{figure}
\centering
\includegraphics[width=0.5\textwidth]
{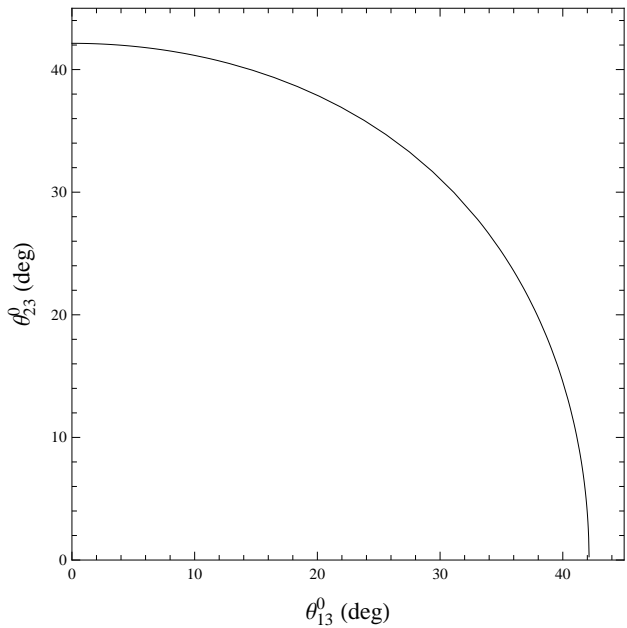}
\caption{Dependence of $\theta_{23}^0$ on $\theta_{13}^0$ for small
  perturbations in the charged lepton sector when $U_\delta^l$ is
  given by Eq.~(\ref{eq:Udl}), and the three mixing angles in the
  PMNS matrix are fixed by the best-fit values of the global fit in
  Table~\ref{tab:data} for the normal hierarchy.}
\label{fg:2}
\end{figure}

\begin{figure}
\centering
\includegraphics[width=0.65\textwidth]
{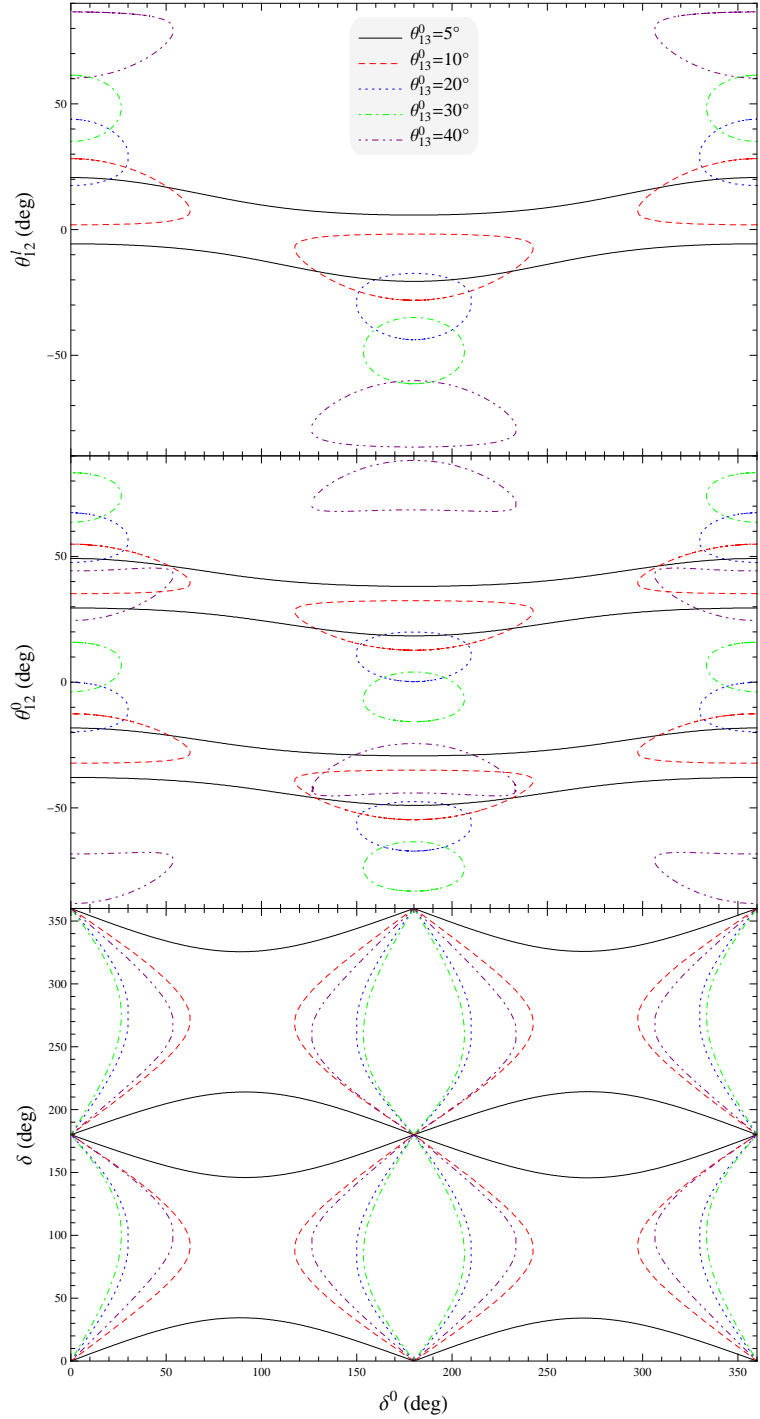}
\caption{Dependence of $\theta_{12}^l$ (top), $\theta_{12}^0$ (middle) and $\delta$
  (bottom) on $\delta^0$ for different values of $\theta_{13}^0$
  for small perturbations in the charged lepton sector when
  $U_\delta^l$ is given by Eq.~(\ref{eq:Udl}), and the three
  mixing angles in the PMNS matrix are fixed by the best-fit values of
  the global fit in Table~\ref{tab:data} for the normal hierarchy.}
\label{fg:3}
\end{figure}

There are eight parameters in Eqs. (\ref{eq:clc23}), (\ref{eq:clc13}) and (\ref{eq:clc12}). We use the best-fit values in Table~\ref{tab:data} for the normal hierarchy to fix $\theta_{12}$, $\theta_{13}$ and $\theta_{23}$. Then for given values of $\theta_{13}^0$ and $\delta^0$, the other three unknown parameters $\theta_{23}^0$, $\theta_{12}^l$ and $\theta_{12}^0$ are determined by the three equations. First, we obtain $\theta_{23}^0$ from Eq.~(\ref{eq:clc23}) for a given $\theta_{13}^0$. Note that the constraints on
$\theta_{23}^0$ and $\theta_{13}^0$ are symmetric for fixed $\theta_{13}$ and $\theta_{23}$, which can be
seen from Fig.~\ref{fg:2}. Then we scan $\theta_{12}^l$ from $[-90^\circ,90^\circ]$ to find solutions to Eq.~(\ref{eq:clc13}) for a given $\delta^0$.
For each solution of $\theta_{12}^l$, we obtain $\theta_{12}^0$ from Eq.~(\ref{eq:clc12}) by scanning $\theta_{12}^0$ from $[-90^\circ,90^\circ]$. Note that we only scan the first and fourth quadrants of 
$\theta_{12}^l$ [$\theta_{12}^0$], because Eq.~(\ref{eq:clc13}) [Eq.~(\ref{eq:clc12})] is only sensitive to the relative sign between the cosine and sine of $\theta_{12}^l$ [$\theta_{12}^0$]. Once we obtain $\theta_{23}^0$, $\theta_{12}^l$ and $\theta_{12}^0$ for given values of $\theta_{13}^0$ and $\delta^0$, the resulting PMNS matrix is completely determined
 (except for the diagonal Majorana phase matrix) from Eqs.~(\ref{eq:Upmns}),~(\ref{eq:Udl}) and~(\ref{eq:U0}). By comparing the PMNS matrix with the standard parametrization, the resulting Dirac phase $\delta$ is also obtained for given values of $\theta_{13}^0$ and $\delta^0$. 
The dependence of $\theta_{12}^l$, $\theta_{12}^0$ and $\delta$ on $\delta^0$ for different values of $\theta_{13}^0$ is shown in Fig.~\ref{fg:3}.
From Figs.~\ref{fg:2} and \ref{fg:3} we see that
the initial mixing angles and the initial Dirac phase can be very different from
their observed values in the PMNS matrix due to small perturbations in the charged lepton
sector.

Generally, perturbations in both the charged lepton and neutrino
sectors will be present. In this case, one must first use the procedure described in this section to find
the corrections to the initial PMNS matrix from perturbations in the
charged lepton sector alone, then use the new PMNS matrix to rotate
to the basis in which the final charged lepton mass matrix is diagonal, 
and ultimately use the procedure described in Section~2 to find the final
corrections to the parameters in the PMNS matrix from perturbations in
the neutrino sector.

\section{Neutrino oscillations with nonstandard interactions}
We now apply our generalized perturbation procedure to a
phenomenological study of neutrino oscillations that are affected by
nonstandard scalar and nonstandard vector interactions
simultaneously. 

As  $\nu_e$ propagate in matter, they scatter on electrons via the V-A interaction. This is described by 
the MSW potential~\cite{Wolfenstein:1977ue}, which is added
to the vacuum oscillation Hamiltonian:
\begin{align}
  \label{eq:H0}
  H=&\frac{1}{2E_\nu}U\begin{pmatrix}
  m_1^2 & 0 & 0 \\
  0 & m_2^2 & 0 \\
  0 & 0 & m_3^2
  \end{pmatrix}U^\dagger+ 
    \begin{pmatrix}
       \sqrt{2} G_F N_e & 0 & 0
      \\
      0 & 0 & 0
      \\
      0 & 0 & 0
    \end{pmatrix}\,,
\end{align}
where $G_F$ is the Fermi constant, $N_e$ is the electron number
density in the medium, and $U$ and $m_i$ are the mixing matrix and
eigenmasses in vacuum, respectively.

New physics beyond the Standard Model can be probed by studying model-independent
nonstandard interactions in neutrino oscillation experiments;
for a recent review see Ref.~\cite{Ohlsson:2012kf}. Most studies of
nonstandard interactions are focused on the vector interaction, which can be described by
effective four-fermion operators of the form
\begin{equation}
  \label{eq:NSI}
  \mathcal{L}_\text{V} =
  \frac{G_{F}}{\sqrt{2}} \epsilon^{V}_{\alpha\beta} \!
        \left[ \bar{\nu}_\alpha \gamma^{\rho} (1 - \gamma^{5}) \nu_\beta \right] \!\!
        \left[ \bar{f} \gamma_{\rho} (1 \pm \gamma^{5}) f \right] + \text{h.c.}\,,
\end{equation}
where $f=u,d,e$ is a charged fermion field, and
$\epsilon_{\alpha\beta}^{V}$ are dimensionless parameters that denote
the strength of the deviation from the standard interactions. Similar to
the MSW term, the matter effect due to the nonstandard vector interaction
modifies the oscillation Hamiltonian by additional potential terms,
$\sqrt{2} G_F N_f \epsilon^{V}_{\alpha\beta}$.

In addition, consider nonstandard scalar interactions, which may arise
from a Lagrangian of the form
\begin{equation}
\mathcal{L}_S=\lambda_\nu^{\alpha\beta}\bar{\nu}_\alpha \nu_\beta\phi+\lambda_f \bar{f}f\phi\,,
\end{equation}
where $\phi$ is a new scalar field, and $\lambda_\nu^{\alpha\beta}$ and
$\lambda_f$ are dimensionless coupling constants for neutrinos and
charged fermions, respectively. In a mean field approximation, the
nonstandard scalar interaction will shift elements of
the neutrino mass matrix by~\cite{Sawyer:1998ac},
\begin{equation}
\epsilon_{\alpha\beta}\approx\frac{\lambda_\nu^{\alpha\beta}}{m_\phi^2}\lambda_fN_f\,,
\end{equation}
where $m_\phi$ is the mass of the scalar field, $N_f$ is the number
density of the charged fermion $f$, which is assumed to be
nonrelativistic.

Tests of the inverse square law of the gravitational force put
stringent $m_\phi$-dependent limits on the coupling of a new scalar field to the nucleon field~\cite{Adelberger:2003zx}.
For $m_\phi$ in the range, $ 10^{-6}{\rm~eV}$ to $10^{-10}{\rm~eV}$,
the current experimental upper limit of $\lambda_N$ varies from $10^{-21}$
to $10^{-22}$~\cite{Schlamminger:2007ht}.
Since
\begin{equation}
\epsilon_{\alpha\beta} \simeq 0.46 \text{ meV } \left(\frac{\lambda_\nu}{10^{-4}}\right) \left(\frac{\lambda_N}{10^{-21}}\right)
\left(\frac{N_f}{N_A/\text{cm}^3}\right)
\left(\frac{10^{-6}\text{ eV}}{m_\phi}\right)^2\,,
\end{equation} 
and $N_f\sim 1 N_A/\text{cm}^3\sim 10^{10} \text{ eV}^3$ on
Earth and $N_f\sim 100 N_A/\text{cm}^3\sim 10^{12} \text{ eV}^3$ in
the solar core where most solar neutrinos are produced, in these environments, a $\lambda_\nu$ of
order $10^{-3}$ gives a mass shift of order 1~meV for $m_\phi=10^{-6} \rm~eV$.
Such $\epsilon_{\alpha\beta}$ values are possible for much smaller values of
$\lambda_\nu$ when $m_\phi < 10^{-6}$~eV.

In the presence of both nonstandard scalar and nonstandard vector interactions, the effective Hamiltonian for neutrino oscillations can be written as
\begin{align}
  \label{eq:H}
  H_{eff} =&\frac{1}{2E_\nu}M_{eff}^\dagger M_{eff} + 
    \sqrt{2} G_F N_e \begin{pmatrix}
       1 & 0 & 0
      \\
      0 & 0 & 0
      \\
      0 & 0 & 0
    \end{pmatrix}+\sqrt{2} G_F N_f
      \begin{pmatrix}
         \epsilon_{ee}^V & \epsilon_{e\mu}^V & \epsilon_{e\tau}^V
        \\
        \epsilon_{e\mu}^{V*} & \epsilon_{\mu\mu}^V & \epsilon_{\mu\tau}^V
        \\
        \epsilon_{e\tau}^{V*} & \epsilon_{\mu\tau}^{V*} & \epsilon_{\tau\tau}^V
      \end{pmatrix}\,,
\end{align}
where the effective mass matrix has the form
\begin{align}
M_{eff}=U^*\begin{pmatrix}
m_1 & 0 & 0 \\
0 & m_2 & 0 \\
0 & 0 & m_3
\end{pmatrix}U^\dagger+\begin{pmatrix}
   \epsilon_{11} & \epsilon_{12} & \epsilon_{13} \\
   \epsilon_{12} & \epsilon_{22} & \epsilon_{23} \\
   \epsilon_{13} & \epsilon_{23} & \epsilon_{33}
   \end{pmatrix}\,.
\end{align}

We apply our generalized perturbation procedure to the study of both
nonstandard scalar and nonstandard vector interactions.  By incorporating the
the corrections to the vacuum oscillation parameters (arising from the scalar interaction)
 into the nonstandard vector interaction
formulas, we immediately obtain new formulas for oscillation
probabilities with both nonstandard scalar and nonstandard vector
interactions. Taking the oscillations of $\nu_\mu$ in long baseline
experiments as an example, the result for the $\nu_\mu$ survival
probability is~\cite{Kopp:2007ne}
\begin{eqnarray}
\label{eq:nsipmm}
 &&P_{\mu\mu} \simeq 1 - s^2_{2\times {23}} \left[ \sin^2 \frac{\Delta m_{31}^2 L}{4E} \right] \nonumber\\
 && - ~
 |\epsilon_{\mu\tau}^V| \cos \phi_{\mu\tau}^V s_{2 \times {23}} \left[ s^2 _{2 \times {23}} (\sqrt{2}G_FN_eL) \sin \frac{\Delta m_{31}^2 L}{2E} + 4  c^2_{2 \times {23}}  \frac{2\sqrt{2}G_FN_eE}{\Delta m_{31}^2} \sin^2 \frac{\Delta m_{31}^2 L}{4E} 
 \right]\nonumber\\
 && +~ (|\epsilon _{\mu\mu}^V| - |\epsilon _{\tau\tau}^V|) s^2_{2 \times {23}} c_{2 \times {23}}\left[  \dfrac{\sqrt{2}G_FN_eL}{2} \sin  \frac{\Delta m_{31}^2 L}{2E} - 2  \frac{2\sqrt{2}G_FN_eE}{\Delta m_{31}^2} \sin^2 \frac{\Delta m_{31}^2 L}{4E}\right]\,,
 \end{eqnarray}
where $\Delta m^2_{31}=m^2_3-m^2_1$, $s_{2\times ij}=\sin 2\theta_{ij}$, $c_{2\times ij}=\cos
2\theta_{ij}$, and $\phi_{\mu\tau}^V=\arg \epsilon_{\mu\tau}^V$. After
replacing $\Delta m_{31}^2\rightarrow \Delta m_{31}^2+2(m_3\delta m_3-m_1\delta m_1)$ and $\theta_{23}\rightarrow \theta_{23}+\delta \theta_{23}$,
where the shifts in $m_i$ and $\theta_{23}$ can be easily obtained from our perturbation results
in Section~2, the new formula for both nonstandard scalar
and nonstandard vector interactions is as follows:
\begin{eqnarray}
\label{eq:pmm}
 &&P_{\mu\mu}  \simeq 1 - s^2_{2\times {23}} \left[ \sin^2 \frac{\Delta m_{31}^2 L}{4E} \right] \nonumber\\
 &&-2\delta\theta_{23}\sin 4\theta_{23} \sin^2 \frac{\Delta m_{31}^2 L}{4E} -\frac{(m_3\delta m_3-m_1\delta m_1)L}{2E}s^2_{2\times {23}}\sin \frac{\Delta m_{31}^2 L}{2E}\nonumber\\
 && - ~
 |\epsilon_{\mu\tau}^V| \cos \phi_{\mu\tau}^V s_{2 \times {23}} \left[ s^2 _{2 \times {23}} (\sqrt{2}G_FN_eL) \sin \frac{\Delta m_{31}^2 L}{2E} + 4  c^2_{2 \times {23}}  \frac{2\sqrt{2}G_FN_eE}{\Delta m_{31}^2} \sin^2 \frac{\Delta m_{31}^2 L}{4E} 
 \right]\nonumber\\
 && +~ (|\epsilon _{\mu\mu}^V| - |\epsilon _{\tau\tau}^V|) s^2_{2 \times {23}} c_{2 \times {23}}\left[  \dfrac{\sqrt{2}G_FN_eL}{2} \sin  \frac{\Delta m_{31}^2 L}{2E} - 2  \frac{2\sqrt{2}G_FN_eE}{\Delta m_{31}^2} \sin^2 \frac{\Delta m_{31}^2 L}{4E}\right]\,.
 \end{eqnarray} 
We see that cancellations between the nonstandard scalar and vector terms are possible, a study of which is beyond the scope of this paper. 

\section{Summary}

We introduced a generalized procedure to study complex
perturbations on Majorana neutrino mass matrices. In the charged lepton basis, we derived
analytic formulas for the corrections to the three mixing angles, and the
Dirac and Majorana phases for arbitrary initial mixing. Since
$m_1$ and $m_2$ are nearly degenerate, the corrections to
$\theta_{12}$ and the Dirac and Majorana phases could be very
large. We performed a numerical analysis on the mass matrices with
$\mu-\tau$ symmetry to illustrate our analytical results, and found that the final Dirac phase can take any
value under small perturbations.

We also studied the scenario in which the charged lepton mass matrix
is not diagonal, and considered perturbations on the charged
lepton mass matrix. We found that small perturbations in the charged
lepton sector give small mixing in the 1-3 and 2-3 sectors, but the
mixing in the 1-2 sector could be potentially large due to the near degeneracy
of $m_e$ and $m_\mu$ (on the scale of $m_\tau$), which could lead to
large corrections to all three mixing angles in the PMNS matrix.

In addition, we showed that using our generalized perturbation procedure, it is straightforward to study
neutrino oscillations with both nonstandard scalar and nonstandard vector interactions.

\vskip 0.1in
{\bf Acknowledgements.}
We are grateful to R.~Sawyer for useful discussions. This research was supported by the DOE under Grant No. DE-SC0010504 and by the Kavli
Institute for Theoretical Physics under NSF Grant No. PHY11-25915.

\appendix

\section{Diagonalization of a $2\times 2$ complex symmetric matrix}
\label{ap:diagonalization}

We show how to diagonalize a $2\times 2$ complex
symmetric matrix
\begin{align}
M=\begin{pmatrix}
a & b \\
b & c \\
\end{pmatrix}=\begin{pmatrix}
|a|e^{i\phi_a} & |b| e^{i\phi_b} \\
|b| e^{i\phi_b} & |c|e^{i\phi_c} \\
\end{pmatrix}\,,
\end{align}
so that
\begin{align}
U^TMU=\begin{pmatrix}
m_1 & 0 \\
0 & m_2 \\
\end{pmatrix}\,,
\end{align}
where $m_1$, $m_2$ are non-negative real numbers, and $U$ is an unitary
matrix.

First, we diagonalize $M$ with a unitary matrix $V$ of the form
\begin{align}
V=\begin{pmatrix}
c_\xi & s_\xi e^{-i\phi} \\
-s_\xi e^{i\phi} & c_\xi \\
\end{pmatrix}\,,
\label{eq:Uform}
\end{align}
where $c_\xi$ and $s_\xi$ denote $\cos \xi$ and $\sin \xi$, respectively:
\begin{align}
V^T MV=\begin{pmatrix}
a c_\xi^2+c s_\xi^2e^{2i\phi}-2bs_\xi c_\xi e^{i\phi} & (ae^{-i\phi}-ce^{i\phi}) s_\xi c_\xi +b(c_\xi^2-s_\xi^2) \\
(ae^{-i\phi}-ce^{i\phi}) s_\xi c_\xi +b(c_\xi^2-s_\xi^2) & a s_\xi^2e^{-2i\phi}+c c_\xi^2+2b s_\xi c_\xi e^{-i\phi} \\
\end{pmatrix}\,.
\end{align}
The diagonalization condition is
\begin{align}
(ae^{-i\phi}-ce^{i\phi}) s_\xi c_\xi +b(c_\xi^2-s_\xi^2)=0\,,
\end{align}
which implies the phase $\phi$ is
\begin{align}
\phi=\arctan \frac{|a|\sin(\phi_a-\phi_b)-|c|\sin(\phi_c-\phi_b)}{|a|\cos(\phi_a-\phi_b)+|c|\cos(\phi_c-\phi_b)}\,,
\end{align}
and the mixing angle $\xi$ is 
\begin{align}
\xi=\frac{1}{2}\arctan\frac{2|b|}{|c|\cos(\phi_c+\phi-\phi_b)-|a|\cos(\phi_a-\phi-\phi_b)}\,.
\end{align}
Also, the two eigenvalues can be written as 
\begin{align}
m_1&=|a c_\xi^2+c s_\xi^2e^{2i\phi}-2bs_\xi c_\xi e^{i\phi}| \,,\nonumber \\
m_2&=|a s_\xi^2e^{-2i\phi}+c c_\xi^2+2b s_\xi c_\xi e^{-i\phi}|\,.
\end{align}
The final unitary matrix that diagonalizes $M$ is
\begin{align}
U=V\begin{pmatrix}
e^{i\omega_1/2} & 0 \\
0 & e^{i\omega_2/2} \\
\end{pmatrix}\,,
\end{align}
where 
\begin{align}
\omega_1&=-\arg \left(a c_\xi^2+c s_\xi^2e^{2i\phi}-2bs_\xi c_\xi e^{i\phi}\right) \,,\nonumber \\
\omega_2&=-\arg \left(a s_\xi^2e^{-2i\phi}+c c_\xi^2+2b s_\xi c_\xi e^{-i\phi}\right)\,.
\end{align}

\section{Right-multiplication}
\label{ap:rtmulti}
We calculate the change of mixing parameters when a
general initial mixing matrix $U_0$ (see Eq.~\ref{eq:U0}) is
multiplied by the following unitary matrix from the right:
\begin{align}
U_{12}(\xi,\phi)=\begin{pmatrix}
c_\xi & s_\xi e^{-i\phi} & 0 \\
-s_\xi e^{i\phi} & c_\xi & 0 \\
0 & 0 & 1\\
\end{pmatrix}\,.
\end{align}
Multiplying $U_0$ on the right by $U_{12}$ yields
\begin{eqnarray}
U &=& R_{23}(\theta_{23}^0)U_{13}(\theta_{13}^0,\delta^0)R_{12}(\theta_{12}^0)
\begin{pmatrix}
   1 & 0 & 0 \\[0.3em]
   0 & e^{i\phi_2^0/2} & 0 \\[0.3em]
   0 & 0 & e^{i\phi_3^0/2}
   \end{pmatrix} U_{12}(\xi,\phi)\nonumber\\
&=& \begin{pmatrix}
   c_{13}^0 \tilde{C}_{12}
 & c_{13}^0 \tilde{S}_{12}^* e^{i\frac{\phi_2^0}{2}}
 & e^{-i(\delta^0-\frac{\phi_3^0}{2})}s_{13}^0
\\ 
   -c_{23}^0 \tilde{S}_{12} - s_{23}^0 s_{13}^0 \tilde{C}_{12} e^{i\delta^0}
 &  (c_{23}^0 \tilde{C}_{12}^* - s_{23}^0 s_{13}^0 \tilde{S}_{12}^*
      e^{i\delta^0})e^{i\frac{\phi_2^0}{2}}
 & e^{i\frac{\phi_3^0}{2}}c_{13}^0s_{23}^0
\\[0.3em]
   s_{23}^0 \tilde{S}_{12} - c_{23}^0 s_{13}^0 \tilde{C}_{12} e^{i\delta^0}
 &(-s_{23}^0 \tilde{C}_{12}^* - c_{23}^0s_{13}^0\tilde{S}_{12}^* e^{i\delta^0})
   e^{i\frac{\phi_2^0}{2}}
 & e^{i\frac{\phi_3^0}{2}}c_{13}^0c_{23}^0
   \end{pmatrix}\,,
\end{eqnarray}
where
\begin{equation}
\tilde{C}_{12} = c_{12}^0 c_\xi - s_{12}^0 s_\xi e^{i\frac{\phi_2^0+2\phi}{2}}
\end{equation}
and
\begin{equation}
\tilde{S}_{12} = s_{12}^0 c_\xi + c_{12}^0 s_\xi e^{i\frac{\phi_2^0+2\phi}{2}}
\end{equation}
are complex. Comparing $U$ to the standard parametrization, we find that 
\begin{equation}
\theta_{23} =\theta_{23}^0\,, \quad \theta_{13}=\theta_{13}^0\,,
\end{equation}
and
\begin{equation}
\theta_{12} = \arcsin(|\tilde {S}_{12}|) =
\arcsin \sqrt{\sin^2(\theta_{12}^0+\xi)
             -\sin(2\theta_{12}^0)\sin(2\xi)\sin^2\frac{\phi_2^0+2\phi}{4}}\,.
\end{equation}

Note that after the right-multiplication the phases of the resulting
mixing matrix are not in the standard form. Defining 
\begin{align}
\alpha &= \rm{arg}(\tilde{C}_{12})
= -\arctan\frac{\tan\theta_{12}^0\tan\xi\sin(\phi_2^0/2+\phi)}
               {1-\tan\theta_{12}^0\tan\xi\cos(\phi_2^0/2+\phi)}\,,
\label{eq:alpha}
\end{align}
\begin{align}
\beta &= \rm{arg}(\tilde{S}_{12})
= \arctan\frac{\tan\xi\sin(\phi_2^0/2+\phi)}
              {\tan\theta_{12}^0+\tan\xi\cos(\phi_2^0/2+\phi)}\,,
\label{eq:beta}
\end{align}
we can write $U$ as
\begin{equation}
U = \begin{pmatrix}
   c_{13}^0 |\tilde{C}_{12}| e^{i\alpha}
 & c_{13}^0 |\tilde{S}_{12}| e^{i(\frac{\phi_2^0}{2}-\beta)}
 & e^{-i(\delta^0-\frac{\phi_3^0}{2})}s_{13}^0
\\
   -c_{23}^0 |\tilde{S}_{12}| e^{i\beta}
   -s_{23}^0 s_{13}^0 |\tilde{C}_{12}| e^{i(\delta^0+\alpha)}
 &( c_{23}^0 |\tilde{C}_{12}|e^{-i\alpha}
   -s_{23}^0 s_{13}^0 |\tilde{S}_{12}|e^{i(\delta^0-\beta)})
      e^{i\frac{\phi_2^0}{2}}
 & e^{i\frac{\phi_3^0}{2}}c_{13}^0s_{23}^0
\\[0.3em]
   s_{23}^0 |\tilde{S}_{12}|e^{i\beta}
  -c_{23}^0 s_{13}^0 |\tilde{C}_{12}| e^{i(\delta^0+\alpha)}
 &(-s_{23}^0|\tilde{C}_{12}|e^{-i\alpha}
   -c_{23}^0s_{13}^0|\tilde{S}_{12}| e^{i(\delta^0-\beta)})
   e^{i\frac{\phi_2^0}{2}}
 & e^{i\frac{\phi_3^0}{2}}c_{13}^0c_{23}^0
   \end{pmatrix}\,.
\end{equation}
On removing the unphysical phases $\phi_e = \alpha$ and $\phi_\mu =
\phi_\tau = \beta$ from the rows, the phases in the second
and third columns match the form of the standard parametrization,
with the Majorana phases shifted by
\begin{align}
\Delta\phi_2 &= -2(\alpha + \beta)\,,
\label{eq:dphi2}
\end{align}
\begin{align}
\Delta\phi_3 &= -2\beta\,,
\label{eq:dphi3}
\end{align}
and the Dirac phase shifted by
\begin{equation}
\Delta\delta = \alpha - \beta\,.
\end{equation}



%
%

\end{document}